\documentclass[12pt]{article}
\usepackage{graphics}
\begin{document}
\title{The size of flavor changing effects\\induced by the symmetry
breaking sector}
\author{\normalsize B. Holdom\thanks{holdom@utcc.utoronto.ca}\\
\small {\em Department of Physics, University of Toronto}\\\small {\em
Toronto, Ontario,} M5S1A7, CANADA}\date{}\maketitle
\begin{picture}(0,0)(0,0)
\put(310,205){UTPT-97-14}
\put(310,190){hep-ph/9706333}
\end{picture}
\begin{abstract} It has recently been shown that strong interactions
underlying electroweak symmetry breaking will induce four-fermion
amplitudes proportional to
\({m_{t}^2}\), which in turn will influence a variety of flavor changing
processes. We argue that the size of these effects are likely to be far
below the current experimental bounds.
\end{abstract}
\baselineskip 19pt

The corrections induced by a new strong sector underlying electroweak
symmetry breaking are conveniently encoded in an effective chiral
Lagrangian. Recently attention has focused on a particular term in this
effective Lagrangian which induces corrections proportional to
\({m_{t}^2}\) in charged current interactions \cite{a}. Integrating out the
\(t\) quark then yields interesting effects in the down-type quark sector,
inducing corrections to \({R_{b}}\) and
\({B_{d}^0}\)--\(\overline{{B_{d}^0}}\) mixing \cite{a}, and various
rare \(B\) and \(K\) decays \cite{b}. All these effects are correlated since
they are related to one parameter in the effective Lagrangian
\cite{b}.

In this note we will provide an estimate of the size of the new parameter,
expressed in terms of the number of new fermion doublets in some
underlying theory of electroweak symmetry breaking. Such an estimate
in the case of the \(S\) parameter proved useful to constrain technicolor
theories. The \(S\) parameter is related to a term in an effective
Lagrangian with coefficient \({L_{10}}= - S/16\pi \) which is completely
analogous to the \({L_{10}^{\it QCD}}\) term appearing at order
\(p^{4}\) in the low energy QCD chiral Lagrangian \cite{e}. Since
\({L_{10}^{\it QCD}}\) is a measured quantity, an estimate for \(S\) is
thereby obtained \cite{c} for QCD-like technicolor theories.

The situation is somewhat different for the parameter of interest here,
which is the coefficient of another term appearing at order \(p^{4}\). In
the QCD case this coefficient is not a measurable quantity since the term
can be removed by the equations of motion. Fortunately in the QCD case
there are quark models which model the chiral symmetry breaking and
which quite successfully reproduce the values of all ten measured
parameters, \({L_{1}}\)--\({L_{10}}\). Such models can then be
expected to provide a reasonable estimate of the new parameter, which
we will refer to as \({L_{11}}\) (corresponding to \({a_{11}}\) in
\cite{a} and \({\alpha _{11}}\) in \cite{b}).

A naive quark model may be based on the nonlinear sigma model where
effects of a single quark loop are considered. This does a fair job of
reproducing those
\({L_{i}}\)'s which happen to correspond to convergent loop integrals
\cite{f}. A more sophisticated quark-based approach leads to the
extraction of all ten parameters \cite{g}. Most convenient for our
purposes is the gauged nonlocal constituent (GNC) quark model \cite{d},
which incorporates the momentum dependence expected for dynamically
generated fermion masses. In QCD the mass function is known to fall as
the square of the momentum (up to logarithms) for large momentum. The
GNC model can incorporate such momentum dependence in an manner
which preserves the local chiral symmetries of the underlying theory. The
mass function then naturally regulates the loop integrals, and successful
values for
\({L_{1}}\)--\({L_{10}}\) are obtained. Here we are interested in
\({L_{11}}\) and its sensitivity to the choice of mass function.

We will focus on the following two terms in the chiral Lagrangian,
\begin{equation}{{\cal{L}}_{\chi }}{\ \ \ni\ \ }{L_{10}} {\rm
Tr}(U^{{\dagger}}{L_{\mu \nu }}UR^{{
\mu \nu }}) + {L_{11}}{\rm Tr}(({{\cal{D}}_{\mu }}V^{\mu })
^{2}){,\label{a}}\end{equation} with
\begin{eqnarray}&&{\it U}(x)=e^{{i\pi (x)^{a}{\sigma _{a}}/v}}{,\ \ 
\ \ }{V_{\mu }}=({D_{\mu }}U)^{{}}U^{{\dagger}}
{,\nonumber}\\&&{D_{\mu }}U={{\partial}_{\mu }}U + i{L_{\mu
 }}U - iU{R_{\mu }}{,\nonumber}\\&&{{\cal{D}}_{\mu }}{V_{\nu
}}={{\partial}_{\mu } }{V_{\nu }} + i[{L_{\mu }}, {V_{\nu
}}]{,\nonumber}\\&&{L_{\mu \nu }}={{\partial}_{\mu }}{L_{\nu }}
 - {{\partial}_{\nu }}{L_{\mu }} - [{L_{\mu }},  {L_{\nu
}}]{,\nonumber}\\&&{L_{\mu }}=({V_{\mu }^a} - {A_{\mu
}^a})\sigma ^{a},{\
\ \ }{R_{\mu }}=({V_{\mu }^a} + {A_{\mu }^a})
\sigma ^{a}{.\nonumber}\end{eqnarray} By identifying \({V_{\mu }^a}
- {A_{\mu }^a}\) with weak gauge fields \(g{{\it W}_{\mu }^a}/2\) and
using the equation of motion for \({{\it W}_{\mu }}\) the \({L_{11}}\)
term can be transformed into a sum of 4-quark operators, which include
the following charged current interaction
\cite{a}.
\begin{equation} - \frac {8{L_{11}}{m_{t}^2}}{v^{4}}\sum _{{\it
i,j}={\it d,s,b}^{}}{V_{{\it ti}}^*}{V_{{ tj}}}{\overline{q}_{{\it
iL}}}{t_{R}}{\overline{t}_{R}}{q _{{\it jL}}}\end{equation}
 \({V_{{\it ti}}}\) are the CKM matrix elements.

The chiral symmetry breaking physics in the underlying theory will
produce finite contributions to the \({L_{10}}\) and \({L_{11}}\)
coefficients. These coefficients become running couplings within the
effective theory, and the finite values we are referring to correspond to
the values renormalized at the chiral symmetry breaking scale. These
values may be determined by choosing some convenient amplitude and
using that to match the effective theory onto the underlying theory (or at
least a model of the underlying theory).

We shall consider the two-point function
\begin{equation}\int e^{{iqx}}{\langle }{V_{\mu a}}(x){V_{\nu b}}(0)
 - {A_{\mu a}}(x){A_{\nu b}}(0){\rangle }dx=i{\Gamma _{\mu \nu
}}(q^{2}){\delta _{{\it ab}}}\end{equation} where the only
contributions from the order \(p^{4}\) chiral Lagrangian are from
\({L_{10}}\) and \({L_{11}}\).
\begin{equation}{\Gamma _{\mu \nu
}^{p^{4}}}({}q^{2})=16{L_{10}} q^{2}{g_{\mu \nu }} +
16({L_{11}} - {L_{10}}){q_{\mu }} {q_{\nu
}}{\label{b}}\end{equation} The naive quark loop yields
\begin{equation}{L_{10}^N}={L_{11}^N}= - \frac {{N_{d}}}{96\pi
^{2}},\end{equation} where \({N_{d}}\) is the number of fermion
doublets contributing in the loop.

The couplings of the gauge fields to the fermions in the GNC model are
the following, where \(\Sigma (p^{2})\) is the dynamical fermion mass
function.
\begin{eqnarray}&&{\Gamma _{V}^{\mu a}}({p_{1}}, q,
{p_{2}}={p_{1}} + q) =i\gamma ^{\mu }\sigma ^{a} - i{\it
G}({p_{2}}, {p_{1}}) ({p_{1}} + {p_{2}})^{\mu }\sigma
^{a}\\&&{\Gamma _{A}^{\mu a}}=i\gamma ^{\mu }{\gamma _{5}}
\sigma ^{a}\\&&{\Gamma _{{\it VV}}^{\mu \nu ab}}({p_{1}},
{q_{1}},  {q_{2}}, {p_{2}}={p_{1}} + {q_{1}} + {q_{2}})= - i[{\it G
}({p_{2}}, {p_{1}})g^{{\mu \nu }}\sigma ^{b}\sigma ^{a}
{\nonumber}\\&&{\ \ \ \ } + \frac {{\it G}({p_{2}}, {p_{1}}) - {\it
G}({p _{1}} + {q_{1}}, {p_{1}})}{({p_{2}} + {p_{1}} +
{q_{1}})^{{}} {\cdot}{q_{2}}}(2{p_{1}} + {q_{1}})^{\mu }({p_{2}}
+ {p_{1}} + {q_{1}})^{\nu }\sigma ^{b}\sigma
^{a}]{\nonumber}\\&&{\ \ \ \ } - i[({q_{1}}, \mu , a)^{{}}{\ 
\leftrightarrow\ }({q_{2}}, \nu , b)^{{}}]\\&&{\Gamma _{{\it
AA}}^{\mu \nu ab}}=0\\&&{\it G}({p_{2}}, {p_{1}})=\frac {\Sigma (
- {p_{2}^2}) - 
\Sigma ( - {p_{1}^2})}{{p_{2}^2} - {p_{1}^2}}\end{eqnarray} Thus
as a consequence of gauge invariance one diagram will have the two
vector fields attached at the same point on the loop. To study the effect of
the momentum dependence in \(\Sigma (p^{2})\) we will consider the
following one parameter family of mass functions satisfying \(\Sigma
(m^{2})=m\).
\begin{equation}\Sigma (p^{2})=\frac {(1 + A)m^{3}}{p^{2} +
Am^{2}}{\label{d}}\end{equation} For \(A{\ \rightarrow\ }\infty \), the
naive quark loop result is obtained, whereas QCD is better modeled by a
value of \(A\) closer to unity.

We plot \({L_{10}^{\it GNC}}/{L_{10}^N}\) and \({L_{11}^{\it
GNC}}/{L_{11}^N}\) in Fig. (1) as a function of \(A\). When \(A=1\)
we see that
\({L_{10}^{\it GNC}}\) is more than twice as large as the naive quark
loop result, which brings it into line with the measured QCD value
\cite{d}.
\({L_{11}^{\it GNC}}\) on the other hand shows surprisingly little
sensitivity to
\(A\), and its value for all \(A\) is close to
\({L_{11}^N}\).\footnote{These results agree with those presented in
\cite{d}, given that the \({L_{11}}\) term in those references is defined
differently so that it appears in (\ref{b}) with the opposite sign.} We also
checked other functional forms for \(\Sigma (p^{2})\) and found that
\({L_{11}^{\it GNC}}\) is generally quite insensitive to the momentum
dependence of the fermion mass function.

In \cite{a} and \cite{b} it was argued that the experimental constraints
from
\({R_{b}}\) and \({B_{d}^0}\)--\(\overline{{B_{d}^0}}\) mixing imply
that \(
\left|  \! {L_{11}} \!  \right|  < .1\). But we now see that this limit is far
above what we could reasonably expect. Our results imply that the limit
is saturated for
\({N_{d}}{\;\approx\;}90\), which is clearly impossible given the
constraint on
\({N_{d}}\) from \(S\). To put it another way, if we assume that a heavy
fourth family (\({N_{d}}=4\)) is still allowed by the constraint on \(S\),
then the contribution to \({L_{11}}\) is only \( - .004\). The shifts in the
various rare \(B\) and \(K\) decay modes from standard model values as
described
\cite{b} would then be less than 5\%.

While our models of strong interactions are admittedly simple-minded,
they strongly suggest that the effects of \({L_{11}}\) will be very
difficult to observe given the constraints on \(S\). We conclude that if
flavor changing effects beyond the standard model are seen, they will
likely have more to do with the physics responsible for quark and lepton
masses rather than the physics responsible for electroweak symmetry
breaking.

\section*{Acknowledgment} This research was supported in part by the
Natural Sciences and Engineering Research Council of Canada.

\begin{figure}
\begin{center} \includegraphics{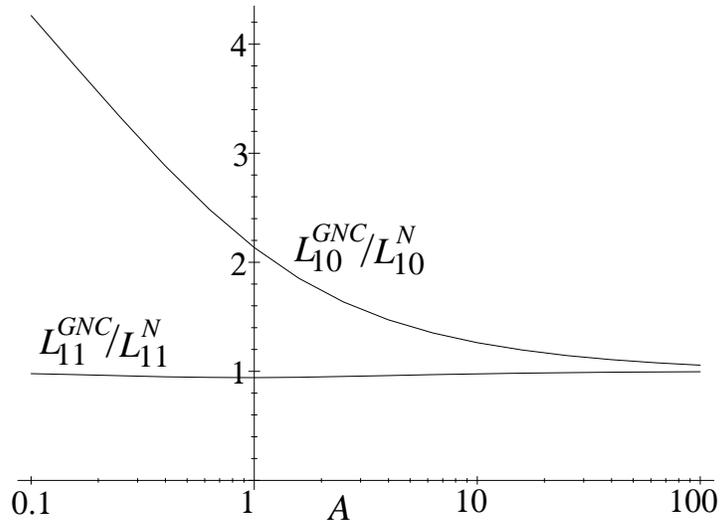}
\end{center}
\caption{\({L_{10}^{\it GNC}}\) and \({L_{11}^{\it GNC}}\) are
coefficients of terms in the chiral Lagrangian (\ref{a}) as determined by
the gauged nonlocal constituent quark model. The naive quark model
gives
\({L_{10}^N}={L_{11}^N}= - {N_{d}}/96\pi ^{2}\). The parameter
\(A\) appears in the fermion mass function in (\ref{d}).}
\end{figure}


\newpage
\begin{thebibliography}{99}
\bibitem{a} D. Comelli, J. Bernabeu, A. Pich, A. Santamaria, Phys. Rev.
Lett.
\textbf{78}, 2902 (1997).
\bibitem{b} G. Burdman, Wisconsin preprint, MADPH-97-996,
hep-ph/9705400.
\bibitem{e} J. Gasser and H. Leutwyler, Nucl. Phys.
\textbf{B250}, 465 (1985).
\bibitem{c} B. Holdom and J. Terning, Phys. Lett.
\textbf{B247}, 88 (1990).
\bibitem{f} e.g. J. Balog, Phys. Lett. \textbf{149B}, 197 (1984).
\bibitem{g} D. Espriu, E. De Rafael, and J. Taron, Nucl. Phys.
\textbf{B345}, 22 (1990), and references therein.
\bibitem{d} B. Holdom, Phys. Rev. \textbf{D45}, 2534 (1992); J.
Terning, Phys. Rev. \textbf{D44}, 887 (1991); B. Holdom, J. Terning,
and K. Verbeek, Phys. Lett.
\textbf{B245}, 612 (1990);
\textbf{B273}, 549E (1991).
\end{thebibliography}
\end{document}